\documentclass[11pt]{article}

\usepackage[margin=1in]{geometry}
\usepackage{graphicx}
\usepackage{hyperref}
\usepackage{caption}
\usepackage{subcaption}
\usepackage{amsmath}
\usepackage{booktabs}
\usepackage{natbib}
\usepackage{float}

\title{Evaluating Workflow Automation Efficiency Using n8n: A Small-Scale Business Case Study}
\author{Ahmed Raza Amir}
\date{}

\begin{document}

\maketitle

\begin{abstract}
Workflow automation has become increasingly accessible through low-code platforms, enabling small organizations and individuals to improve operational efficiency without extensive software development expertise. This study evaluates the performance impact of workflow automation using n8n through a small-scale business case study. A representative lead-processing workflow was implemented to automatically store data, send email confirmations, and generate real-time notifications. Experimental benchmarking was conducted by comparing 20 manual executions with 25 automated executions under controlled conditions. The results demonstrate a significant reduction in the average execution time from 185.35 seconds (manual) to 1.23 seconds (automated), corresponding to an approximately 151× reduction in execution time. Additionally, manual execution exhibited an error rate of 5\%\,while automated execution achieved zero observed errors. The findings highlight the effectiveness of low-code automation in improving efficiency, reliability, and operational consistency for small-scale workflows.
\end{abstract}

\section{Introduction}
Digital transformation has accelerated the adoption of automation technologies in all industries, enabling organizations to streamline repetitive tasks, reduce operational errors, and improve service responsiveness. Traditionally, automation solutions required specialized programming skills and substantial development effort, which limited accessibility to small businesses and individual practitioners. The emergence of low-code and no-code platforms has significantly lowered this barrier by allowing users to design workflows through visual interfaces and reusable components.

Workflow automation tools such as n8n enable rapid integration between data sources, communication platforms, and business services with minimal coding. These platforms offer an attractive solution for small-scale operations where efficiency gains can directly impact productivity and customer satisfaction. Despite growing adoption, quantitative evaluations of low-code automation performance in small-business contexts remain limited.

This study investigates the efficiency of workflow automation using n8n through a practical case study involving automated lead processing. The workflow captures customer input, stores records in a cloud database, sends confirmation emails, and delivers real-time notifications. Performance is evaluated by comparing manual execution against automated execution in terms of execution time, error rate, and execution stability.

The primary objective of this research is to quantify the operational benefits of low-code automation under controlled conditions and provide empirical evidence to support its effectiveness for small-scale applications. The findings aim to inform practitioners, educators, and early-stage developers about the practical advantages and limitations of adopting workflow automation technologies.

\section{Related Work}
Previous studies have examined the role of low-code and no-code platforms in improving development speed, accessibility, and operational efficiency. \citep{khankhoje2022beyond} provide a comparative analysis of traditional automation, low-code, and no-code approaches, highlighting trade-offs between flexibility, development speed, and user accessibility. Their findings emphasize the growing relevance of visual automation tools for rapid solution development.

\citep{viswanadhapalli2025future} explores the impact of low-code and no-code platforms on intelligent automation and AI decision making, demonstrating how such platforms democratize advanced technologies by reducing development complexity and cost barriers. Similarly, \citep{ilesanmi2025nocode} reviews real-world business applications of no-code AI, illustrating practical efficiency gains across customer service, marketing, and operational workflows.

\citep{jangam2024scalability} discusses the scalability and performance limitations of low-code and no-code platforms in large enterprise deployments, suggesting hybrid approaches to address performance constraints. \citep{ajiga2024automation} and \citep{dalsaniya2022intelligent} further examine how software automation and intelligent automation enhance operational efficiency by reducing errors and improving process consistency.

While existing literature establishes the benefits of automation and low-code platforms at the conceptual and enterprise levels, few studies provide quantitative experimental measurements in small-scale, practitioner-driven environments. This study contributes empirical performance data through controlled benchmarking of manual versus automated execution in a lightweight workflow automation scenario.

\section{System Design}
\begin{figure}[H]
\centering
\includegraphics[width=0.99\linewidth]{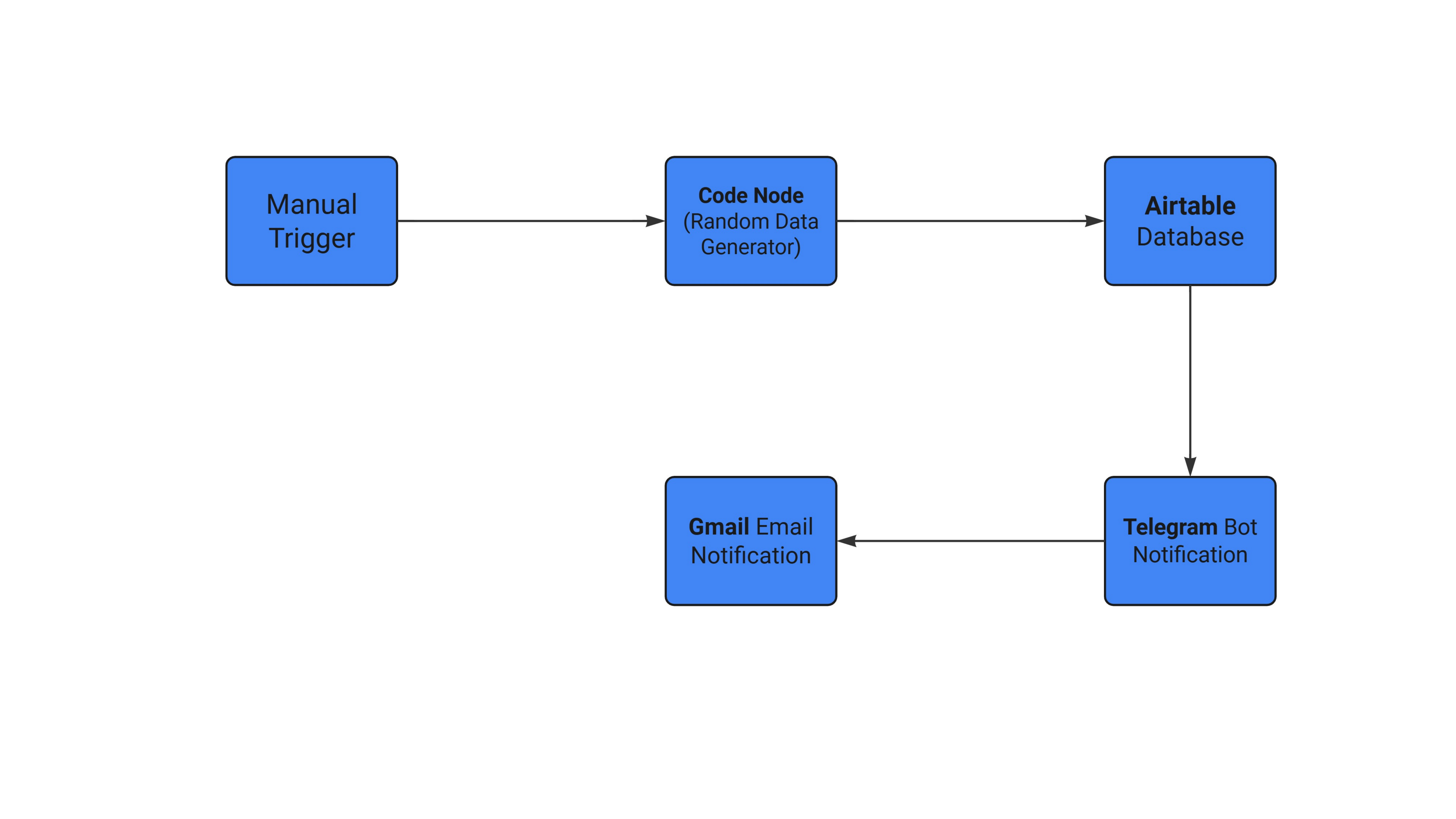}
\caption{Workflow architecture implemented in n8n.}
\label{fig:workflow_architecture}
\end{figure}

The workflow was implemented using n8n Cloud for workflow orchestration and Airtable as the cloud-based data storage platform. The final workflow executes a linear sequence consisting of a manual trigger, code node, Airtable database storage, Gmail email notification, and Telegram bot notification. The system architecture follows a linear workflow consisting of a manual trigger, a code node, and an Airtable record creation node. The code node generates randomized sample lead data, including customer name, email, and message fields, along with experimental metadata such as a unique run identifier, execution source label, and timestamp. Each workflow execution inserts a single structured record into Airtable, enabling systematic data collection and traceability. Figure~\ref{fig:workflow_architecture} illustrates the implemented workflow architecture and notification pipeline.

Stored data fields include RunID, Source, ExecutedAt, Name, Email, and Message. All automated executions are labeled as automated to enable a direct comparison with manually executed workflows. Execution timestamps are generated programmatically using the ISO 8601 format to ensure consistency and reproducibility across runs.

During initial configuration and testing, the workflow was executed multiple times to validate data integrity, field consistency, and system stability. Early configuration issues related to credential setup and node reconnection were resolved prior to experimental benchmarking. Subsequent validation runs demonstrated reliable behavior with no missing fields or execution failures.

Airtable was selected due to its structured schema, stable API integration with n8n, and ease of exporting data for downstream analysis. The system was later extended to include automated email confirmation and Telegram notification modules, enabling end-to-end process automation for realistic operational evaluation.

\begin{table}[H]
\centering
\caption{Experimental environment and system configuration.}
\label{tab:environment}
\begin{tabular}{ll}
\hline
\textbf{Component} & \textbf{Description} \\
\hline
Workflow platform & n8n Cloud \\
Database & Airtable \\
Email service & Gmail SMTP \\
Notification service & Telegram Bot API \\
Execution mode & Manual trigger and automated execution \\
Measurement method & Stopwatch and n8n execution logs \\
Total experimental runs & 45 runs \\
\hline
\end{tabular}
\end{table}

\section{Experimental Methodology}
The experimental evaluation was conducted to compare manual workflow execution with automated execution using the implemented n8n workflow. Two experimental modes were defined: manual execution and automated execution. In the manual mode, the workflow steps were performed by a human operator, including manual data entry into the database, sending confirmation emails, and sending notification messages. Execution time measurement began when data entry was initiated and ended when all required actions were completed. Operational mistakes or delays were recorded as execution errors.

A total of 20 manual executions were performed. Execution time for each run was measured using a stopwatch and recorded in seconds. One execution error was observed during manual testing, resulting in an error rate of five percent. Execution time variability was recorded to capture consistency and human performance variation.

In the automated mode, the workflow was executed using the n8n automation pipeline without manual intervention. A total of 25 automated executions were performed. Execution duration was obtained directly from the n8n execution history logs, which provide precise workflow runtime measurements. Email and messaging delivery latency were not separately benchmarked and are therefore not included in the automated timing measurements. No execution failures were observed during automated testing.

All experimental data were recorded in structured spreadsheets for analysis. Summary statistics including average execution time, minimum and maximum execution time, and error rate were computed to enable quantitative comparison between manual and automated execution modes.

\section{Results}
This section presents the experimental comparison between manual workflow execution and automated execution using n8n. The evaluation focuses on execution time, error rate, and execution stability.

\begin{figure}[H]
    \centering
    \includegraphics[width=0.85\linewidth]{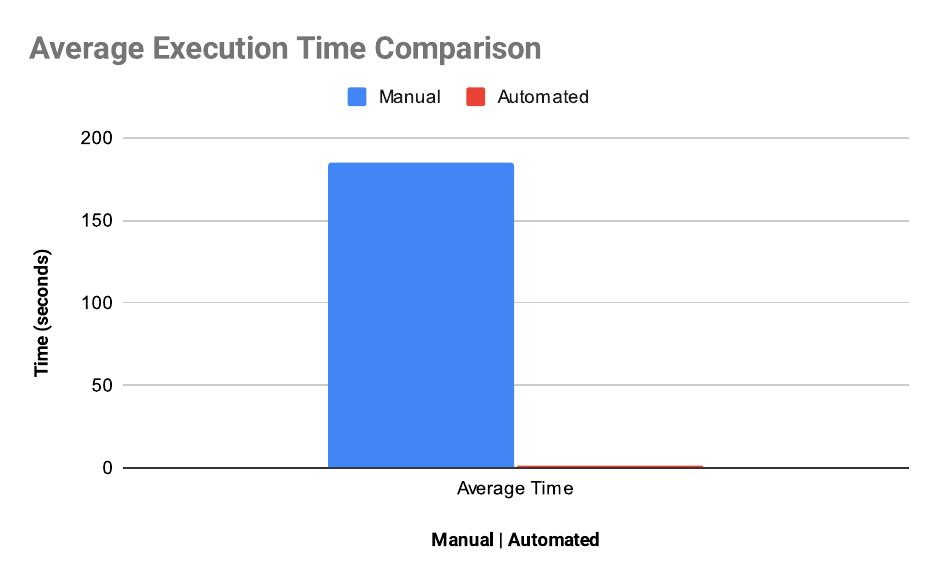}
    \caption{Average execution time comparison between manual and automated workflow execution.}
    \label{fig:avg_time}
\end{figure}

A total of 20 manual runs and 25 automated runs were recorded. The average execution time for manual processing was 185.35 seconds, while the automated workflow achieved an average execution time of 1.23 seconds. This represents an approximate 151 times speed improvement when using automation. Figure~\ref{fig:avg_time} illustrates the substantial reduction in execution time achieved through automation. \begin{table}[H]
\centering
\caption{Performance comparison between manual and automated workflow execution.}
\label{tab:performance_summary}
\begin{tabular}{lcc}
\hline
\textbf{Metric} & \textbf{Manual Execution} & \textbf{Automated Execution} \\
\hline
Number of runs & 20 & 25 \\
Average time (seconds) & 185.35 & 1.23 \\
Minimum time (seconds) & 146 & 1.03 \\
Maximum time (seconds) & 266 & 2.79 \\
Error count & 1 & 0 \\
Error rate & 5\%\ & 0\%\ \\
\hline
\end{tabular}
\end{table}

\begin{figure}[H]
    \centering
    \includegraphics[width=0.75\linewidth]{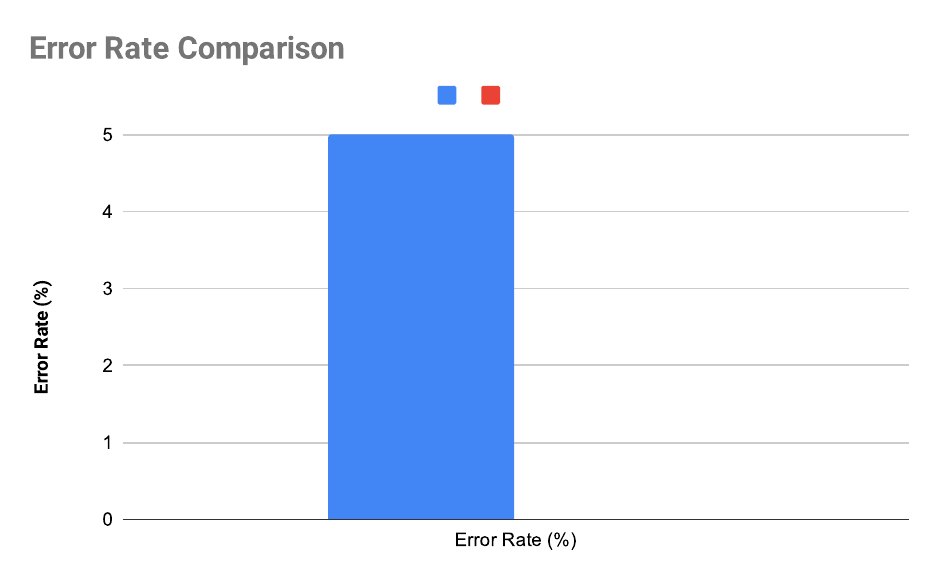}
    \caption{Error rate comparison between manual and automated workflow execution.}
    \label{fig:error_rate}
\end{figure}

Manual execution resulted in one error across 20 runs, corresponding to an error rate of five percent. In contrast, the automated workflow completed all 25 executions without any failures, yielding an error rate of zero percent. Figure~\ref{fig:error_rate} highlights the reduction of human-induced errors through automation.

\begin{figure}[H]
    \centering
    \includegraphics[width=0.85\linewidth]{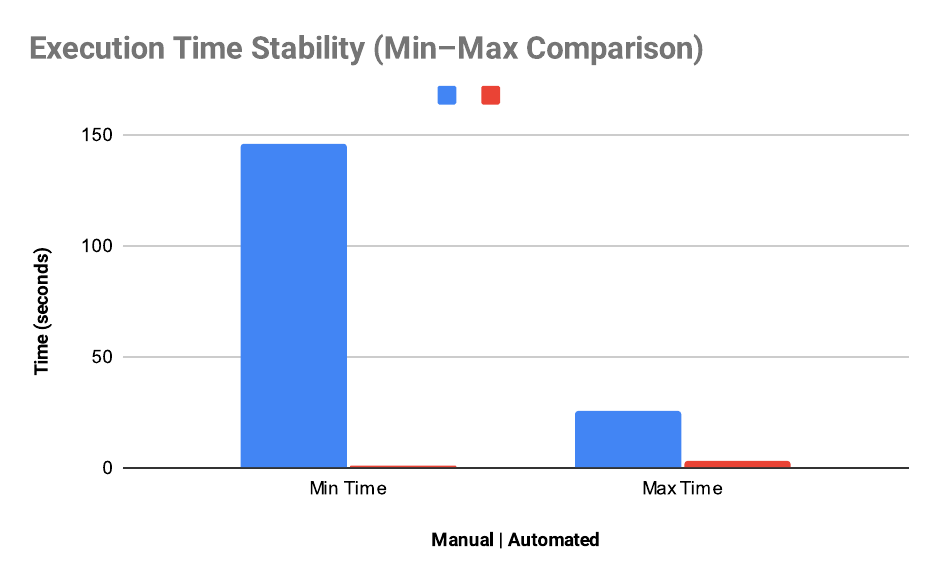}
    \caption{Execution time stability comparison showing minimum and maximum execution times for manual and automated workflows.}
    \label{fig:stability}
\end{figure}

Execution stability was evaluated by analyzing the variation between minimum and maximum execution times. Manual execution exhibited a minimum time of 146 seconds and a maximum time of 266 seconds, indicating high variability. Automated execution demonstrated significantly lower variation, with execution times ranging from 1.03 seconds to 2.79 seconds. Figure~\ref{fig:stability} illustrates the improved stability and predictability of automated execution.

Overall, the results confirm that low-code workflow automation significantly improves operational efficiency, reduces human error, and provides consistent execution performance in small-scale workflow scenarios.

\section{Discussion}
The experimental results demonstrate that low-code workflow automation using n8n can significantly enhance operational efficiency in small-scale business scenarios. The observed 151 times reduction in execution time highlights the productivity gains achievable through automation for repetitive and rule-based tasks. Such improvements translate into faster response times, reduced operational workload, and improved service reliability.

The elimination of human errors during automated execution further emphasizes the reliability advantages of automation. Manual workflows are susceptible to typing mistakes, missed steps, and inconsistencies caused by fatigue or multitasking. Automation enforces consistent logic execution, ensuring uniform outcomes across repeated runs.

Execution stability is another important observation. Manual execution exhibited wide variability in completion time, whereas automated execution remained within a narrow time range. This predictability is valuable for operational planning, scaling, and monitoring in real-world deployments.

Overall, the findings support the practical adoption of low-code automation platforms for lightweight workflows and provide empirical evidence of measurable benefits in efficiency, reliability, and consistency.

\section{Limitations}
Although the experimental results demonstrate strong performance improvements, several limitations should be acknowledged. First, the workflow was triggered manually to maintain controlled experimental conditions. While this supports repeatability, it does not fully represent event-driven environments where workflows may be triggered by external systems such as web forms, webhooks, or APIs, and where network latency and service delays can affect performance.

Second, the dataset size was limited to 20 manual runs and 25 automated runs. While sufficient to show clear trends, larger datasets would provide stronger statistical confidence and enable deeper variability analysis.

Third, the evaluation was conducted using an n8n cloud environment and a single Airtable configuration. Performance characteristics may differ under alternative infrastructure, concurrent execution loads, or different storage backends.

Finally, the study focused primarily on execution time, error rate, and stability. Other important factors such as security considerations, credential management, long-term maintainability, and total cost of ownership were not evaluated in this experiment.

\section{Future Work}
Future research can extend this work in several directions. One improvement is to evaluate event-driven triggers such as webhooks, form submissions, or API-based integrations to better simulate real operational environments and assess the effect of external latency.

Scalability testing under concurrent execution loads would provide insight into performance and reliability when handling higher volumes, which is important for production adoption. Comparative evaluations across multiple low-code automation platforms could further contextualize performance and usability trade-offs.

Additional work may also include security and reliability analysis, including fault tolerance, retry strategies, monitoring, and recovery mechanisms. Finally, integrating analytics or AI-based components could enable intelligent decision-making workflows that go beyond deterministic rule-based automation.

\section{Conclusion}

This study evaluated the performance impact of low-code workflow automation using n8n through a small-scale business case study. Experimental benchmarking demonstrated a substantial reduction in execution time from manual processing to automated execution, achieving more than a 150 times speed improvement. Automation also eliminated observed human errors and provided consistent and predictable execution behavior.

The results highlight the practical value of low-code automation platforms for improving operational efficiency, reliability, and scalability in lightweight workflows. By combining empirical measurements with a reproducible implementation, this work provides actionable evidence supporting the adoption of workflow automation by small organizations and individual practitioners.

Future extensions may include event-driven triggers, scalability testing under concurrent workloads, and integration with intelligent decision-making components. Overall, the findings confirm that low-code automation offers a viable and impactful pathway toward digital transformation in resource-constrained environments.

\bibliography{references}
\end{document}